# Discovery of an Intermediate Nematic State in a Bilayer Kagome Metal ScV$_6$Sn$_6$

*Camron Farhang,[1] William R. Meier,[2] Weihang Lu,[1] Jiangxu Li,[3] Yudong Wu,[1] Shirin Mozaffari,[2] Richa P. Madhogaria,[2] Yang Zhang,[3,4] David Mandrus,[2,5] Jing Xia[1]*

[1] *Department of Physics and Astronomy, University of California, Irvine, Irvine, CA 92697, USA*

[2] *Department of Materials Sciences and Engineering, University of Tennessee-Knoxville, Knoxville, Tennessee 37996, USA*

[3] *Department of Physics and Astronomy, University of Tennessee-Knoxville, Knoxville, Tennessee 37996, USA*

[4] *Min H. Kao Department of Electrical Engineering and Computer Science, University of Tennessee-Knoxville, Knoxville, Tennessee 37996, USA*

[5] *Materials Science and Technology Division, Oak Ridge National Laboratory, Oak Ridge, TN 37831, USA*

**Nematicity, where rotational symmetry of the crystal lattice is spontaneously broken, is a ubiquitous phenomenon in correlated quantum matter, often intertwining with other orders to produce a richer spectrum of phases. Here we report a new phase transition in high-quality ScV$_6$Sn$_6$ bilayer kagome metal at a temperature T$^*$, occurring seven Kelvins below the charge density wave (CDW) transition at T$_{CDW}$, as indicated by thermodynamic, transport, and optical measurements. This emerging intermediate phase does not exhibit spontaneous time-reversal-symmetry breaking, as evidenced by zero-field Sagnac interferometer experiments. However, it displays a strong, spontaneous (strain- and field-free) anisotropy in the kagome plane between T$^*$ and T$_{CDW}$, as revealed by transport and optical polarization rotation measurements. Additionally, a pronounced depolarization effect detected by the Sagnac interferometer further confirms its nematic nature. This intermediate nematic phase, alongside the recently discovered intra-unit cell nematic order at much lower temperatures, presents a diverse landscape of nematicities at multiple length and temperature scales, distinguishing it from those observed in kagome metals AV$_3$Sb$_5$. Our findings highlight ScV$_6$Sn$_6$ and the broader RM$_6$X$_6$ intermetallic family as fertile platforms for realizing symmetry-breaking phases driven by a unique interplay of competing CDW instabilities, kagome physics, and Van Hove singularities.**

Spontaneous symmetry breaking underlines a multitude of fundamental natural phenomena, from the mass acquisition of elementary particles [1] to various solid-state matters in nature [2]. In the context of quantum materials, the breaking of lattice rotational symmetry, or nematicity, plays a crucial role in some of the most intriguing systems, including high-temperature cuprate superconductors [3], quantum Hall systems [4], and, more recently, layered kagome metals [5–9]. The vanadium kagome metal $ScV_6Sn_6$ [10], with V atoms forming kagome bilayers (Fig.1d), belongs to the extensive $RT_6X_6$ family of intermetallics (R = rare earth; T = V, Cr, Mn, Fe, Co; X = Ge, Sn). This family exhibits complex magnetism and nontrivial topological electronic properties [11–17]. Its band structure [18] features multiple Van Hove singularities (VHS) near the Fermi level, a Dirac point at the K point, and a flat band above the Fermi level $E_F$, suggesting the potential for novel correlated phases. Below $T_{CDW} \sim 80$ to $95\ K$ [10,18–23], $ScV_6Sn_6$ hosts a $\sqrt{3} \times \sqrt{3} \times 3$ CDW state tripling the unit cell along the c-axis [10] as depicted in Fig.4b, accompanied by a shift of VHS away from $E_F$ [18]. Additionally, a recent STM study has revealed an intra-unit-cell nematic order and the associated Fermi surface deformation that occur at a much lower temperature of $70\ K$ [9]. Compared to the $AV_3Sb_5$ (A = K, Rb, Cs) family of kagome metals [24], which host both CDW [24] and nematic orders [5–8], $ScV_6Sn_6$ exhibits unique CDW instabilities due to the smallness of Sc atoms [25,26], giving rise to competing CDW fluctuations [19,20,27]. Understanding if these instabilities lead to new CDW-driven nematicities [6] is of key importance. Here we report transport, thermodynamics, optical reflectivity, polarization rotation, and Sagnac interferometry measurements in high-quality $ScV_6Sn_6$ crystals, revealing a unusual nematic phase emerging below $T_{CDW}$ and persisting over a narrow temperature range. Upon further cooling, in-plane rotational symmetry is restored as competing charge-density wave fluctuations subside, distinguishing this intermediate nematic state (Fig.4a) from any previously reported nematic phase.

**Emergence of an intermediate phase inside the CDW state**

High-quality $ScV_6Sn_6$ samples were synthesized using the tin flux method [10]. They are nicely faceted metallic hexagonal blocks about $1\ mm$ in size with optically flat ab-plane surfaces (Fig.1a inset). The ab-plane resistivity $\rho$ during cooldown is plotted in Fig.1a. The CDW transition is evident as a sharp drop in in-plane resistivity on cooling through $T_{CDW} \approx 93\ K$, consistent with previous reports [10,18–23]. Below $T_{CDW}$, there is a pronounced second resistivity drop at $T^* \approx 86\ K$, which has only been clearly observed in a $ScV_6Sn_6$ crystal with similar growth conditions to this study [18], while hints of this second drop appear as a long tail in temperature-dependent resistivity in other reports [19,23]. The room-temperature resistivity of $ScV_6Sn_6$ samples used in this study is $90\ \mu\Omega \cdot cm$, yielding a residual resistivity ratio (RRR) of 9, comparable to reference [18] but about twice as large as other reports [10,18–23]. The improved RRR suggests higher crystal quality and may explain why the $T^*$ transition is only clearly observed in this work and in [18].

The drop in resistivity alone doesn't guarantee a phase transition. A thermodynamic signature, often considered a "smoking gun", has never been reported for a second phase transition in $ScV_6Sn_6$ [10,18–23]. To elucidate whether $T^*$ marks a true phase transition, we have performed specific heat $C_p$ measurements on the same sample during cooldown, as shown in Fig.1c. A sharp main peak in $C_p$ centered at 93 $K$ signals the expected CDW transition [10,19,22]. Additionally, a new peak in $C_p$ emerges at 86 $K$, coinciding with the lower temperature ab-plane resistivity jump, providing a smoking gun for a phase transition at $T^*$.

It is crucial to rule out the trivial scenario in which the $T^*$ transition arises merely from a sample region with a reduced $T_{CDW}$. To address this, we performed local optical reflectivity measurements at 1.55 $\mu m$ wavelength during cooldown at numerous focused spots spread over the ab-plane surface. One such measurement is presented in Fig.1b: the optical reflectivity drops from 0.648 to 0.644 when the sample is cooled below $T_{CDW}$, and it drops further from 0.644 to 0.642 at $T^*$. Similar reflectivity drops at both $T_{CDW}$ and $T^*$ were observed at other spots (Extended Data Fig.1), confirming the local nature of the "double transition". In a metal that is isotropic in the ab-plane, a reduced optical reflectivity usually indicates a decreased plasma frequency and hence less numerous carriers, consistent with what was found in Hall effect measurements on a similar sample [18]. The increased carrier mobility [18] due to a decreased electronic scattering by CDW fluctuations [20] accounts for the lower resistivity in the CDW phase, despite diminished density of states.

**Testing spontaneous time-reversal symmetry breaking (TRSB)**

Beyond the lattice-translational symmetry breaking characteristic of the CDW phase, spontaneous TRSB in ScV$_6$Sn$_6$ has been reported as hidden magnetism in a muon spin relaxation ($\mu SR$) study [22]. To investigate whether TRSB is responsible for the $T^*$ transition, we employ a zero-area Sagnac interferometer [28] that exclusively detects magneto-optic Kerr (MOKE) signals arising from microscopic TRSB and rejects any non-TRSB effects such as anisotropy [29–31]. This is because the sourcing aperture for one light is also the receiving aperture for the other time-reversed counterpropagating light. Consequently, Onsager's relations guarantee zero signal in the absence of microscopic TRSB. Spatial imaging and temperature-dependent measurements of spontaneous MOKE signals $\theta_K$ were performed on the ab-plane surface using a microscope version of the Sagnac interferometer [32–35]. To align the chiralities of possible TRSB domains, a training magnetic field is applied during cooldown and subsequently removed at the lowest temperature. Fig.2(a) depicts a spontaneous $\theta_K$ image acquired at $T = 3\ K$, taken after removing a 0.3 $T$ magnetic field during cooldown. No discernable spontaneous MOKE signal was observed in the image. To further search for any changes in the signal across $T^*$ with the highest resolution, we perform zero field warmups (ZFW) at numerous locations with different training fields. One example of ZFW after $B = +0.3\ T$ training is shown in Fig.2b, and another example after $B = -0.3\ T$ training is presented in Fig.2c. Despite these efforts, we didn't observe any onset of $\theta_K$ at either $T_{CDW}$ or $T^*$ with an uncertainty of $\pm 0.01\ \mu rad$. In fact, across the whole temperature range, there is no discernable onset at any temperature with an uncertainty of $\pm 0.02\ \mu rad$. Additional ZFW measurements are presented in Extended Data Fig.3, further reinforcing the conclusion that there is no evidence for TRSB at any temperature in our sample, which stands in stark contrast to the $\mu SR$ study [22]. It is noteworthy that compared to our ScV$_6$Sn$_6$ crystal, the CDW transition in the $\mu SR$ study occurs at a significantly lower temperature of $80\ K$ [22] compared to most other studies [10,18–23], suggesting a presence of impurities that may account for the observed hidden magnetism in their sample.

In the absence of spontaneous TRSB in our sample, the MOKE signal in a magnetic field originates from Pauli paramagnetism [10,18] in ScV$_6$Sn$_6$, which lacks strongly magnetic atoms. Fig.2(b) presents a MOKE image obtained at $T = 3\ K$ in a magnetic field of $-0.3\ T$, revealing a small but exceptionally uniform $\theta_K$ signal of $-14.0\ \mu rad$ with a spatial variance of only $0.1\ \mu rad$ across the entire ab-plane surface. Temperature-dependent $\theta_K$ measurements (Fig.2e and Extended Data Fig.2a for $B = +0.3\ T$ cooldowns, Fig.2f and Extended Data Fig.2b for $B = -0.3\ T$ cooldowns) reveal a 10% increase in $\theta_K$ at $T_{CDW}$. While this initially appears to contradict the concurrent reduction in carrier density [18], calculations of spin split bands under a magnetic field revealed a significant enhancement of spin polarization in the CDW phase, leading to larger MOKE signals. Fig.2h and Fig.2i show the calculated band structure and density of states of ScV$_6$Sn$_6$ under an external magnetic field $B$ for the high-temperature and CDW phases respectively, with spin-up and spin-down channels marked in red and blue respectively. In the high-temperature phase, spin polarization at the fermi level is 0.10/cell, increasing to 0.19/cell in the CDW phase, explaining the $\theta_K$ enhancement despite the reduced density of states.

A more intriguing observation is the absence of any change in $\theta_K$ across $T^*$, echoing the lack of any discernible kink in the magnetic susceptibility $\chi$ at $T^*$ in a similar crystal [18]. Since Pauli paramagnetism arises from the imbalance in spin-polarized carriers, this decoupling between the $T^*$ transition and the magnetic channel implies that the $T^*$ transition is likely associated to a rearrangement of charge carriers rather than a change in their total quantity.

**Nematicity in the intermediate state**

One possible charge rearrangement is nematicity [4] that spontaneously breaks rotational symmetry. In ScV$_6$Sn$_6$, a recent STM study revealed an intra-unit-cell nematic order accompanied by Fermi surface deformation [9]. The intra-unit-cell nematic order emerges at a lower temperature ($70\ K$) and in a different atomic layer than the CDW order, suggesting a weaker correlation with the CDW order [9]. In contrast, our directional resistance and optical polarization rotation measurements below reveal a distinct type of nematic order in ScV$_6$Sn$_6$ that is strongly linked to $T_{CDW}$ and $T^*$.

Zero-magnetic-field in-plane transport anisotropy is revealed in directional transport, as shown in Fig. 3a. A second ScV$_6$Sn$_6$ crystal was carefully cut into a $500\ \mu m$ square in the ab-plane with $200\ \mu m$ thickness along the c-axis, as shown in the inset of Fig. 3a. Four $25\ \mu m$-diameter gold wires were connected to the four corners to measure resistances $R_{xx} \equiv \frac{V_x}{I_x}$ and $R_{yy} \equiv \frac{V_y}{I_y}$, where the in-plane x and y directions are defined as along and perpendicular to the a-axis of the kagome lattice, respectively. Above $T_{CDW}$, $R_{xx}$ and $R_{yy}$ closely track each other, with a small offset (factor of 1.127) attributed to minor misalignments of electrical contacts. Therefore, we plot $1.127R_{xx}$ and $R_{yy}$ in Fig. 3a, showing near-perfect overlap when cooling down from 150 K to $T_{CDW}$. However, below $T_{CDW}$, a pronounced anisotropy develops, which diminishes but doesn't completely vanish below $T^*$. To better visualize the temperature dependence of the in-plane transport anisotropy, we define $\frac{R_{yy}}{1.127R_{xx}}$ as a phenomenological anisotropy parameter and plot it in Fig.3b. Above $T_{CDW}$, this ratio remains within $1 \pm 0.01$, confirming rotational symmetry in the kagome plane. At $T_{CDW}$, a sudden anisotropy suddenly appears as $\frac{R_{yy}}{1.127R_{xx}}$ drops to 0.84 within a fraction of a Kelvin. Upon further cooling past $T^*$, the ratio recovers to 0.90, signaling a rapid but incomplete recovery of rotational symmetry.

To locally probe the in-plane anisotropy and to determine its principal axes, we performed optical polarization rotation measurements [7,32,36] on the first $ScV_6Sn_6$ crystal. Optical rotation ($\theta_T$) at 1.55 $\mu m$ wavelength was measured under normal incidence onto the ab-plane while varying the incident polarization angle $\alpha$. Fig.3c presents polar plots of $\theta_T(\alpha)$ at various temperatures, with the inset depicting the sample and crystallographic axes. Between ~85 K and ~95 K, $\theta_T$ exhibits both sign and magnitude variations, reaching ±0.5 $mrad$ in a two-fold symmetry pattern described by $\theta_T = \theta_P \sin(2\alpha - A)$. This clearly indicates spontaneous breaking of the six-fold in-plane rotational symmetry. A fitted parameter $A \sim 30°$ suggests that the principal anisotropy axis (green lines in Fig.3c) aligns with the a-axis of the kagome lattice. Above $T_{CDW}$, $\theta_T$ is negligible, indicating preserved rotational symmetry. Below $T^*$, its behavior becomes more complex, displaying significant spatial variations. Extended Data Fig.4 presents $\theta_T$ measured at multiple locations: at point R, $\theta_T$ diminishes below $T^*$, whereas at point M and L, the two-fold symmetry pattern persists down to 70 K and 60 K, respectively. This inhomogeneity may explain why resistance anisotropy decreases but doesn't fully disappear below $T^*$.

Local nematicity also manifests itself in the depolarization effects detected by the Sagnac interferometer. As described in Methods, linear and circular birefringence can cause depolarization to the reflected circularly polarized light beams. Consequently, in a zero-area Sagnac interferometer, some reflected light loses coherence and does not participate in the interference, appearing as a drop in the coherent reflection $P_2$ (and $P_0$ and $P_1$). Such dips are clearly visible between $T_{CDW}$ and $T^*$ in $P_2$ (and $P_0$) traces presented in Extended Data Fig.5, indicating strong depolarization effects in this temperature range. In contrast, the MOKE signal $\theta_K$ is calculated as the ratio between $P_1$ and $P_2$, and is not influenced by depolarization effects. To ascertain that the observed dip in $P_2$ is not due to a change in optical reflectivity, we compare optical reflectivity (red, Fig.3d) and the coherent reflection $P_2$ (blue, Fig.3e). Notably, these two signals are uncorrelated between $T_{CDW}$ and $T^*$ indicating strong depolarization effects, which is evident in Fig.4a where we plot optical depolarization (red line) as the coherent reflection $P_2$ (Fig.3d) divided by reflectivity (Fig.3e) normalized to 80 $K$. Similar to polarization rotation, spatial variations are observed in the magnitude of the $P_2$ dip, ranging from 3% to 1% at various locations (Extended Data Fig.5).

**Discussion**

In Fig.4a, we summarize the nematic orders in ScV$_6$Sn$_6$ by plotting the scaled anisotropy observed in various local measurements. The anisotropic polarization rotation (yellow) represents fitted $\theta_P$ values from Fig.3c using $\theta_T = \theta_P \sin(2\alpha - A)$, while the optical depolarization (red) is calculated as the coherent reflection $P_2$ (Fig.3d) divided by reflectivity (Fig.3e), both highlighting the presence of the intermediate anisotropic phase between $T^*$ and $T_{CDW}$ in high quality crystals. The intermediate phase is well separated in temperature from the recently discovered intra-cell nematic order below 70 $K$ in a STM study [9], represented in Fig.4a by the anisotropic Bragg peaks (blue) calculated as the ratio between Bragg peak intensities at $q_3$ and $q_2$ (adopted from [9]). Unlike clear signatures observed at $T^*$ for the intermediate phase, no anomalies were detected in transport, optical, or specific heat measurements for the intra-cell nematic order. This suggests that the latter is a much more subtle phenomenon, likely of a different origin.

These findings parallel the existence of two distinct temperatures associated with two-fold anisotropy in the CDW state of single layer kagome metals AV$_3$Sb$_5$ [24]. A broad range of experiments indicate the breaking of the underlying $C_6$ rotational symmetry [5–7,37,38] either at $T_{CDW}$ or at a much lower temperature $T'$. Initial nematic susceptibility measurements in CsV$_3$Sb$_5$ reported a diverging anisotropic (E$_{2g}$) elastoresistivity response below $T'$ [6], serving as evidence for electronic nematicity, that is, the partial melting of CDW that restores translational symmetry but breaks rotational symmetry [4]. However, recent more careful experiments [39,40] on CsV$_3$Sb$_5$ reported the absence of the E$_{2g}$ signal, suggesting that the $T'$ transition is not nematic in nature. Even more intriguing, a recent transport study [41] on CsV$_3$Sb$_5$ found that while unperturbed samples show no transport anisotropy, the application of either a magnetic field or strain induced a pronounced transport anisotropy below $T'$. These complex and seemingly contradictory findings suggest the coexistence of two distinct nematic states below $T'$ in AV$_3$Sb$_5$, making it difficult to disentangle them. In contrast, the clear separation between intra-cell nematic below 70 $K$ [9] and the intermediate nematic state in ScV$_6$Sn$_6$ allows easier independent study, which is further facilitated in the intermediate phase by the transport, optical and thermodynamic signatures without any external magnetic field or strain. Future nematic susceptibility measurements [39] will be crucial in determining whether these two states arise from electronic nematic instability.

Another likely origin of the unusual intermediate nematic phase in ScV$_6$Sn$_6$ is the unique competition between different CDW instabilities [19,20,27]. Typically, materials tend to break symmetry upon cooling, resulting in states of lower symmetries. However, counter examples exist, such as the quantum Hall stripe to isotropic phase transition from 100 mK to 20 mK at half-filled high Landau levels restoring in-plane rotational symmetry [42], and the ferromagnetic to antiferromagnetic transition during cooling in intermetallic magnetic compound FeRh restoring time-reversal symmetry [43–45]. In ScV$_6$Sn$_6$ varied CDW phases have been predicted [25] and the lattice degrees of freedom have been shown experimentally to play a fundamental role in establishing the $\sqrt{3} \times \sqrt{3} \times 3$ CDW ground state [19,20,27]. In a nesting-driven CDW arising from Peierls instability, the diverging electronic susceptibility and phonon-softening occur at the same wavevector $\vec{q}$, and cause a static CDW at $q_s$. In a typical electron-phonon-coupling (EPC) driven CDW [46], the strength of EPC peaks at $\overrightarrow{q_{EPC}}$ in the vicinity of phonon-softening regions, resulting in a static CDW at $\overrightarrow{q_{EPC}}$. Conversely, in ScV$_6$Sn$_6$ inelastic X-ray scattering found that EPC peaks at $q_s = (\frac{1}{3}, \frac{1}{3}, \frac{1}{3})$ corresponding to the observed static $\sqrt{3} \times \sqrt{3} \times 3$ CDW, while the softest phonon modes occur at $q^* = (\frac{1}{3}, \frac{1}{3}, \frac{1}{2})$ corresponding to a competing short-range $\sqrt{3} \times \sqrt{3} \times 2$ CDW [20]. The dominance of the $q^*$ mode at higher temperatures is surpassed by the $q_s$ mode below $T_{CDW}$, leading to the long-range $q_s$ CDW state. The proximity of $T^*$ to $T_{CDW}$ suggests that the observed nematicity is closely linked to the lattice degrees of freedom. It is plausible that remnants of the competing $q^*$ mode slightly below $T_{CDW}$ further lower the rotational symmetry, giving rise to the observed nematicity. At a lower temperature as the competing $q^*$ mode subsides, rotational symmetry is restored. In our ScV$_6$Sn$_6$ sample this process must take place at a well-defined temperature $T^*$ in a sizable fraction of the bulk volume giving rise to the observed signatures in specific heat and resistance, while it persists below $T^*$ at other locations as detected in local optical anisotropy measurements (Extended Data Fig.4), resulting in some remnant resistance anisotropy below $T^*$ (Fig.3b).

One possible scenario considers the stacking degree of freedom. Below $T_{CDW}$, the static $q_s$ CDW state preserves in-plane $C_6$ rotational symmetry within each kagome layer, as illustrated in Fig.4b, where only Vanadium atoms are shown for clarity, though Sc and Sn atoms also undergo shifts. However, in the presence of the competing $q^*$ CDW, perfect interlayer phase matching may be softened, allowing a finite interlayer phase shift. In the case of CsV$_3$Sb$_5$ recent Raman measurements [47] have proposed a $\pi$ phase interlayer shift being responsible for the observed $C_6$ symmetry breaking in the $q = \left(\frac{1}{2}, \frac{1}{2}, \frac{1}{2}\right)$ CDW. For the $q_s = (\frac{1}{3}, \frac{1}{3}, \frac{1}{3})$ CDW in bilayer kagome metal ScV$_6$Sn$_6$, this would correspond to an inter-bilayer $\frac{2\pi}{3}$ phase shift. Fig.4c illustrates such a staggered configuration where $C_6$ symmetry is broken due to a misalignment of the $C_6$ centers, which may be restored at lower temperatures once the $q^*$ mode is suppressed to enforce perfectly stacked layers.

Alternatively, the mixing of $q_s$ and $q^*$ CDWs could directly lead to anisotropy. Although $q_s$ and $q^*$ share the same in-plane component $(\frac{1}{3}, \frac{1}{3})$, their relative phase varies across kagome layers resulting in anisotropy in some layers. For example, in Fig.4d, we plot the lattice distortion due to mixing of these two CDWs at height $z = 2.25\ c$. A relative phase shift of $\Delta\phi = \frac{3\pi}{4}$ results in an anisotropic state marked by the dashed green line. The $C_6$ symmetry is restored at lower temperatures when the $q^*$ mode diminishes.

In summary, ScV$_6$Sn$_6$ presents a diverse landscape of two distinct and well-separated nematicities at different length and temperature scales. This is distinct from the nematic state reported in kagome superconductors AV$_3$Sb$_5$ [5–8]. Our results highlight ScV$_6$Sn$_6$ and the family of RM$_6$X$_6$ intermetallics as a fertile platform for realizing symmetry-breaking phases arising from a unique combination of competing CDW instabilities, kagome physics, and Van Hove singularities.

**Methods:**

**Crysal Growth:** ScV$_6$Sn$_6$ samples were synthesized using the tin flux method, starting with an atomic ratio of Sc/Lu:V:Sn = 1:6:60. Sc/Lu pieces (Alfa Aesar, 99.9%), V pieces (Alfa Aesar, 99.8%), and Sn shot (Alfa Aesar, 99.9999%) were placed into an alumina Canfield crucible set. The crucible assembly was sealed in a fused silica ampule, heated to 1150 °C over 12 hours, maintained at this temperature for 15 hours, and then cooled to 780 °C over 300 hours. At 780 °C, the flux was separated from the crystals by inverting the ampule and centrifuging. This process produced hexagonal metallic crystals with typical lateral dimensions ranging from 0.5 to 4 mm. The elemental composition and approximate stoichiometry of the structure were characterized and confirmed via scanning electron microscopy (SEM), energy-dispersive spectroscopy (EDS), and powder x-ray diffraction (XRD). The detailed results of these characterizations are presented in [10,18].

**Specific Heat** measurements were carried out using the heat capacity option of the Physical Property Measurement System (PPMS) by Quantum Design with the crystal mounted with Apiezon N grease. Specific heat data were collected by PPMS during cooldown by applying heat pulses and analyzing the subsequent cooling trace.

**Electrical Transport:** Sample resistivity and anisotropic resistances were measured using a resistance bridge. Electric contacts were made with 25 $\mu m$-diameter gold wires that are connected to the four corners of the samples (Fig.1a inset) using a home-made micro-spot-welding setup. This results in small contact area ($< 50 \ \mu m$) and low contact resistance ($< 0.1 \ m\Omega$).

**Optical Reflectivity** is measured in normal reflection of a 1.55 $\mu m$ wavelength optical beam that is focused onto a single spot on the ab-plane surface. The recorded reflected optical power is compared with that from a reference sample (a highly reflective mirror) to obtain the optical reflectivity.

**Sagnac MOKE measurements** are performed using a zero-loop fiber-optic Sagnac interferometer as shown in Fig.2g. The beam of light is routed by a fiber circulator to a fiber polarizer. After the polarizer the polarization of the beam is at 45° to the axis of a fiber-coupled electro-optic modulator (EOM), which generates 4.6 MHz time-varying phase shifts $\phi_m \sin(\omega t)$, where the amplitude $\phi_m = 0.92$ rad between the two orthogonal polarizations that are then launched into the fast and slow axes of a polarization maintaining (PM) single-mode fiber. Upon exiting the fiber, the two orthogonally polarized linearly polarized beams are converted into right- and left-circularly polarizations by a quarter-wave plate (QWP) and are then focused onto the sample. After reflection from the sample, the same QWP converts the reflected beams back into linear polarizations with exchanged polarization axes. The two beams then pass through the PM fiber and EOM but with exchanged polarization modes in the fiber and the EOM. At this point, the two beams have gone through the same path but in opposite directions, except for a phase difference of $\Delta\varphi$ from reflection off the magnetic sample and another time-varying phase difference by the modulation of EOM. This nonreciprocal phase shift $\Delta\varphi$ between the two counterpropagating circularly polarized beams upon reflection from the sample is twice the Kerr rotation $\Delta\varphi = 2\theta_K$. The two beams are once again combined at the detector and interfere to produce an optical signal $P(t)$:

$$P(t) = \frac{1}{2} P[1 + \cos(\Delta\varphi + \phi_m \sin(\omega t))]$$

, where P is the returned power if the modulation by the EOM is turned off. For MOKE signals that are slower that the 4.6 MHz modulation frequency used in this experiment, we can treat $\Delta\varphi$ as a slowly time-varying quantity. And $P(t)$ can be further expanded into Fourier series with the first few orders listed below:

$$\begin{aligned}
P(t)/P = &\frac{1}{2}[1 + J_0(2\phi_m)] \\
&+ (\sin(\Delta\varphi) J_1(2\phi_m)) \sin(\omega t) \\
&+ (\cos(\Delta\varphi) J_2(2\phi_m)) \cos(2\omega t) \\
&+ 2 J_3(2\phi_m)) \sin(3\omega t) \\
&+ \cdots
\end{aligned}$$

, where $J_1(2\phi_m)$ and $J_2(2\phi_m)$ are Bessel J-functions. Lock-in detection was used to measure the first three Fourier components: the average (DC) power ($P_0$), the first harmonics ($P_1$), and the second harmonics ($P_2$). And the Kerr rotation can then be extracted using the following formula:

$$\theta_K = \frac{1}{2}\Delta\varphi = \frac{1}{2}\tan^{-1}\left[\frac{J_2(2\phi_m)P_1}{J_1(2\phi_m)P_2}\right]$$

**Sagnac Measurement of Depolarization Effects:** The above calculations of the Sagnac interferometer assume absence of any depolarization effects in the sample, i.e. after the right- and left-circularly polarized are reflected from the sample surface, they will remain in perfect circular polarizations. In the presence of linear or circular birefringence and dichroism of the sample, however, these depolarization effects will not guarantee preserving perfect circular polarizations. And the two reflected beams, after passing the quarter-wave plate again, become elliptical instead of being perfectly linearly polarized causing the leakage of some optical power into the "wrong" polarization axes of the optical fiber. And a small fraction of the light will thus be incoherent with the major beams and won't participate in the interference. The remaining coherent parts of optical powers need to be multiplied by a correction factor of $(1 - D)$, where $D$ is a small number quantifying the depolarization effects from the sample. For example, the second harmonics $P_2$ (denoted as coherent reflection in the main text and in Fig.3e) is changed from $(\cos(\Delta\varphi) J_2(2\phi)_m) \cos(2\omega t) P$ to $(1-D)(\cos(\Delta\varphi) J_2(2\phi)_m) \cos(2\omega t) P$, showing up as a "drop" in $P_2$. Since a decrease in optical reflectivity can cause a drop in the reflected total power $P$ and hence a drop in $P_2$ even without any depolarization effects, it is necessary to compare measured coherent reflection $P_2$ and optical reflectivity to determine whether the observed "drop" in $P_2$ is indeed caused by depolarization.

We note that the MOKE signal ($\theta_K = \frac{1}{2} tan^{-1} \left[\frac{J_2(2\phi_m)P_1}{J_1(2\phi_m)P_2}\right]$ is calculated using the ratio of $P_1$ and $P_2$, both of which experience the same reduction in the presence of depolarization effects. Therefore, the depolarization effects will not affect the $\theta_K$ measurement.

**Polarization Rotation:** As shown in Fig.3f, the beam of light is routed through a free-space polarizer to produce a linearly polarized beam. A polarization-independent beam splitter (half mirror) transmits half of the beam and reflects the other half, which is discarded. The transmitted beam passes through a pinhole and then a half-wave plate (HWP), which is mechanically rotated such that its principal fast axis is at an angle $\frac{\alpha}{2}$ to the polarization direction of the beam. The resulting beam after the HWP has its polarization direction rotated by angle $\alpha$. The beam then passes through the optical window of the cryostat and gets reflected by the sample. The returned light beam passing the same pinhole a second time is in general elliptical with the major axis rotated by the total polarization rotation $\alpha + \theta_T$. After passing through the HWP a second time, its polarization direction is rotated by $-\alpha$, and becomes $\theta_T$. And the same polarization-independent beam splitter reflects half the returned beam towards a Wollaston prism, which separates and directs two orthogonal polarizations s and p toward two balanced detectors. The recorded powers of s and p-polarization components are P1 and P2 respectively. The Wollaston prism is rotated at a $\frac{\pi}{4}$ angle such that with a gold mirror calibration sample P1 and P2 are "balanced": $\Delta P = P1 - P2 \sim 0$. The optical amplitudes $E1$ and $E2$ at detectors 1 and 2 are:

$$E1 = E0 \cos\left(\frac{\pi}{4} - \theta_T\right)$$

$$E2 = E0 \cos\left(\frac{\pi}{4} + \theta_T\right)$$

, where $E0$ is the total amplitude. Since optical intensity $I = E^2$, the sum and difference of the two intensities $I1$ and $I2$ are:

$$I1 + I2 = E1^2 + E2^2 = E0^2 \cos^2\left(\frac{\pi}{4} + \theta_T\right) + E0^2 \cos^2\left(\frac{\pi}{4} - \theta_T\right) = E0^2$$

$$I1 - I2 = E1^2 - E2^2 = E0^2 \cos^2\left(\frac{\pi}{4} - \theta_T\right) - E0^2 \cos^2\left(\frac{\pi}{4} + \theta_T\right) = E0^2 \sin(2\theta_T)$$

Hence:

$$\frac{I1 - I2}{I1 + I2} = \sin(2\theta_T)$$

As optical power is proportional to intensity $P \propto I$, we can extract $\theta_T$ as:

$$\theta_T = \frac{1}{2}\arcsin\left(\frac{I1 - I2}{I1 + I2}\right) = \frac{1}{2}\arcsin\left(\frac{P1 - P2}{P1 + P2}\right) = \frac{1}{2}\arcsin\left(\frac{\Delta P}{P1 + P2}\right)$$

**Density functional theory (DFT) Calculations**: All ab initio calculations were performed using the Vienna Ab-initio Simulation Package (VASP) based on density functional theory (DFT). The exchange-correlation interactions were treated using the generalized gradient approximation (GGA) within the Perdew-Burke-Ernzerhof (PBE) functional. Brillouin zone integration was carried out using a Monkhorst-Pack k-point mesh of 10x10x10. The plane-wave energy cutoff was set to 500 eV for expanding the wavefunctions. For the high temperature phase, which belongs to the P6/mmm space group, the optimized lattice constants were $a = b = 5.467$Å and $c = 9.16$ Å. For the $\sqrt{3} \times \sqrt{3} \times 3$ charge density wave (CDW) phase, which belongs to the R32 space group, the calculated lattice constants were $a = b = 9.46$ Å and $c = 27.41$ Å. Spin-polarized calculations were performed to account for the Zeeman effect induced by the external magnetic field. The spin-up and spin-down components of the electronic structure were analyzed separately to extract the spin polarization at the Fermi level. An external magnetic field was introduced via the $B_{ext}$ (BEXT) parameter in VASP. The applied magnetic field induces Zeeman splitting of the electronic states, and the magnitude of the field was controlled by adjusting the value of BEXT (in eV).

**Acknowledgements**

This project was supported by NSF award DMR-2419425 and the Gordon and Betty Moore Foundation EPiQS Initiative, Grant # GBMF10276. W.R.M. and D.M. acknowledge support from the Gordon and Betty Moore Foundation's EPiQS Initiative, Grant GBMF9069 awarded to D.M.. S.M. and R.P.M. acknowledge the support from AFOSR MURI (Novel Light-Matter Interactions in Topologically Non-Trivial Weyl Semimetal Structures and Systems), grant# FA9550-20-1-0322. J. L. and Y. Z. were supported by the National Science Foundation Materials Research Science and Engineering Center program through the UT Knoxville Center for Advanced Materials and Manufacturing (DMR-2309083).

**Author Contributions**
J.X. conceived and supervised the project. C.F., D.L., Y.W., and J.X. carried out the polarization rotation, Sagnac, reflectivity, specific heat and transport measurements. S.M., W.R.M., R.P.M., and D.M. grew the crystals. J.L. and Y.Z. performed the spin polarization calculations. J.X. drafted the paper with the input from all authors. All authors contributed to the discussion of the manuscript.

**Competing interests**

The authors declare no competing interest.

**Materials & Correspondence**

Correspondence and requests for materials should be addressed to Jing Xia (xia.jing@uci.edu).

**References:**

1. Higgs, P. W. Broken Symmetries and the Masses of Gauge Bosons. *Phys. Rev. Lett.* **13**, 508–509 (1964).

2. Anderson, P. W. More Is Different: Broken symmetry and the nature of the hierarchical structure of science. *Science* **177**, 393–396 (1972).

3. Fradkin, E., Kivelson, S. A. & Tranquada, J. M. Colloquium: Theory of intertwined orders in high temperature superconductors. *Rev. Mod. Phys.* **87**, 457–482 (2015).

4. Fradkin, E., Kivelson, S. A., Lawler, M. J., Eisenstein, J. P. & Mackenzie, A. P. Nematic Fermi Fluids in Condensed Matter Physics. *Annual Review of Condensed Matter Physics, Vol 1* **1**, 153–178 (2010).

5. Xiang, Y. *et al.* Twofold symmetry of c-axis resistivity in topological kagome superconductor CsV3Sb5 with in-plane rotating magnetic field. *Nat Commun* **12**, 6727 (2021).

6. Nie, L. *et al.* Charge-density-wave-driven electronic nematicity in a kagome superconductor. *Nature* **604**, 59–64 (2022).

7. Wu, Q. *et al.* Simultaneous formation of two-fold rotation symmetry with charge order in the kagome superconductor CsV3Sb5 by optical polarization rotation measurement. *Phys. Rev. B* **106**, 205109 (2022).

8. Zhao, H. *et al.* Cascade of correlated electron states in the kagome superconductor CsV3Sb5. *Nature* **599**, 216–221 (2021).

9. Jiang, Y.-X. *et al.* Van Hove annihilation and nematic instability on a kagome lattice. *Nat. Mater.* **23**, 1214–1221 (2024).

10. Arachchige, H. W. S. *et al.* Charge Density Wave in Kagome Lattice Intermetallic ScV6Sn6. *Phys. Rev. Lett.* **129**, 216402 (2022).

11. Madhogaria, R. P. *et al.* Topological Nernst and topological thermal Hall effect in rare-earth kagome ScMn 6 Sn 6. *Phys. Rev. B* **108**, 125114 (2023).

12. Ghimire, N. J. *et al.* Competing magnetic phases and fluctuation-driven scalar spin chirality in the kagome metal YMn6Sn6. *Science Advances* **6**, eabe2680 (2020).

13. Pokharel, G. *et al.* Electronic properties of the topological kagome metals YV6Sn6 and GdV6Sn6. *Phys. Rev. B* **104**, 235139 (2021).

14. Pokharel, G. *et al.* Highly anisotropic magnetism in the vanadium-based kagome metal TbV6Sn6. *Phys. Rev. Materials* **6**, 104202 (2022).

15. Hu, Y. *et al.* Tunable topological Dirac surface states and van Hove singularities in kagome metal GdV6Sn6. *Sci. Adv.* **8**, eadd2024 (2022).

16. Gu, Y. *et al.* Phonon mixing in the charge density wave state of ScV6Sn6. *npj Quantum Mater.* **8**, 1–8 (2023).

17. Ortiz, B. R. *et al.* Stability frontiers in the AM6X6 kagome metals; The LnNb6Sn6 (Ln:Ce–Lu,Y) family and density-wave transition in LuNb6Sn6. Preprint at https://doi.org/10.48550/arXiv.2411.10635 (2024).

18. Mozaffari, S. *et al.* Universal sublinear resistivity in vanadium kagome materials hosting charge density waves. *Phys. Rev. B* **110**, 035135 (2024).

19. Hu, Y. *et al.* Phonon promoted charge density wave in topological kagome metal ScV6Sn6. *Nat Commun* **15**, 1658 (2024).

20. Cao, S. *et al.* Competing charge-density wave instabilities in the kagome metal ScV6Sn6. *Nat Commun* **14**, 7671 (2023).

21. DeStefano, J. M. *et al.* Pseudogap behavior in charge density wave kagome material ScV6Sn6 revealed by magnetotransport measurements. *npj Quantum Mater.* **8**, 1–7 (2023).

22. Guguchia, Z. *et al.* Hidden magnetism uncovered in a charge ordered bilayer kagome material ScV6Sn6. *Nat Commun* **14**, 7796 (2023).

23. Hu, T. *et al.* Optical spectroscopy and band structure calculations of the structural phase transition in the vanadium-based kagome metal ScV6Sn6. *Phys. Rev. B* **107**, 165119 (2023).

24. Ortiz, B. R. *et al.* Z2 Topological Kagome Metal with a Superconducting Ground State. *Phys. Rev. Lett.* **125**, 247002 (2020).

25. Tan, H. & Yan, B. Abundant Lattice Instability in Kagome Metal ScV6Sn 6. *Phys. Rev. Lett.* **130**, 266402 (2023).

26. Meier, W. R. *et al.* Tiny Sc allows the chains to rattle: Impact of Lu and Y doping on the charge density wave in ScV6Sn6. *J. Am. Chem. Soc.* **145**, 20943–20950 (2023).

27. Korshunov, A. *et al.* Softening of a flat phonon mode in the kagome ScV6Sn6. *Nat Commun* **14**, 6646 (2023).

28. Xia, J., Beyersdorf, P. T., Fejer, M. M. & Kapitulnik, A. Modified Sagnac interferometer for high-sensitivity magneto-optic measurements at cryogenic temperatures. *Applied Physics Letters* **89**, 062508 (2006).

29. Xia, J., Maeno, Y., Beyersdorf, P. T., Fejer, M. M. & Kapitulnik, A. High resolution polar Kerr effect measurements of Sr2RuO4: Evidence for broken time-reversal symmetry in the superconducting state. *Physical Review Letters* **97**, 167002 (2006).

30. Schemm, E. R., Gannon, W. J., Wishne, C. M., Halperin, W. P. & Kapitulnik, A. Observation of broken time-reversal symmetry in the heavy-fermion superconductor UPt3. *Science* **345**, 190–193 (2014).

31. Xia, J. *et al.* Polar Kerr-effect measurements of the high-temperature YBa2Cu3O6+x superconductor: Evidence for broken symmetry near the pseudogap temperature. *Physical Review Letters* **100**, (2008).

32. Farhang, C., Wang, J., Ortiz, B. R., Wilson, S. D. & Xia, J. Unconventional specular optical rotation in the charge ordered state of Kagome metal CsV3Sb5. *Nat Commun* **14**, 5326 (2023).

33. Gong, C. *et al.* Discovery of intrinsic ferromagnetism in two-dimensional van der Waals crystals. *Nature* **546**, 265–269 (2017).

34. Thomas, S. *et al.* Localized Control of Curie Temperature in Perovskite Oxide Film by Capping-Layer-Induced Octahedral Distortion. *Phys. Rev. Lett.* **119**, 177203 (2017).

35. Farhang, C. *et al.* Revealing the Origin of Time-Reversal Symmetry Breaking in Fe-Chalcogenide Superconductor FeTe1–xSex. *Phys. Rev. Lett.* **130**, 046702 (2023).

36. Xu, Y. *et al.* Three-state nematicity and magneto-optical Kerr effect in the charge density waves in kagome superconductors. *Nat. Phys.* (2022) doi:10.1038/s41567-022-01805-7.

37. Xu, Y. *et al.* Three-state nematicity and magneto-optical Kerr effect in the charge density waves in kagome superconductors. *Nat. Phys.* 1470–1475 (2022) doi:10.1038/s41567-022-01805-7.

38. Li, H. *et al.* Rotation symmetry breaking in the normal state of a kagome superconductor KV3Sb5. *Nat. Phys.* **18**, 265–270 (2022).

39. Liu, Z. *et al.* Absence of E2g Nematic Instability and Dominant A1g Response in the Kagome Metal CsV3Sb5. *Phys. Rev. X* **14**, 031015 (2024).

40. Frachet, M. *et al.* Colossal *c*-Axis Response and Lack of Rotational Symmetry Breaking within the Kagome Planes of the CsV3⊠Sb5 Superconductor. *Phys. Rev. Lett.* **132**, 186001 (2024).

41. Guo, C. *et al.* Correlated order at the tipping point in the kagome metal CsV3Sb5. *Nat. Phys.* **20**, 579–584 (2024).

42. Fu, X. *et al.* Anomalous Nematic States in High Half-Filled Landau Levels. *Phys. Rev. Lett.* **124**, 067601 (2020).

43. Lewis, L. H., Marrows, C. H. & Langridge, S. Coupled magnetic, structural, and electronic phase transitions in FeRh. *J. Phys. D: Appl. Phys.* **49**, 323002 (2016).

44. Hamara, D. *et al.* Ultra-high spin emission from antiferromagnetic FeRh. *Nat Commun* **15**, 4958 (2024).

45. Li, G. *et al.* Ultrafast kinetics of the antiferromagnetic-ferromagnetic phase transition in FeRh. *Nat Commun* **13**, 2998 (2022).

46. Zhu, X., Guo, J., Zhang, J. & Plummer, E. W. Misconceptions associated with the origin of charge density waves. *Advances in Physics: X* **2**, 622–640 (2017).

47. Jin, F. *et al.* π Phase Interlayer Shift and Stacking Fault in the Kagome Superconductor $CsV_3Sb_5$. *Phys. Rev. Lett.* **132**, 066501 (2024).

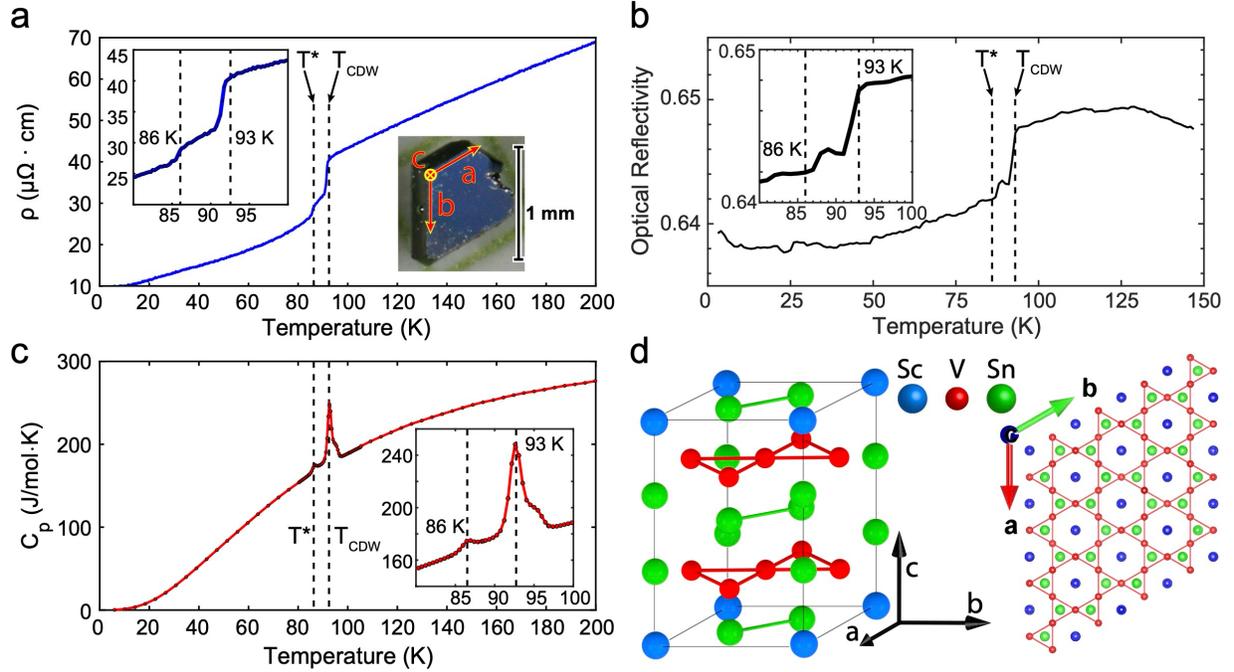

**Figure 1. Emergence of an intermediate phase between $T^*$ and $T_{CDW}$ inside the CDW state of ScV$_6$Sn$_6$.**
**(a)** ab-plane resistivity $\rho$ showing sharp drops when the sample is cooled through $T_{CDW}$ and $T^*$. Left inset: resistivity near transition temperatures. Right inset: image of the measured ScV$_6$Sn$_6$ single crystal sample. **(b)** Optical reflectivity measured with the 1.55 $\mu m$ optical beam focused onto a single spot on the ab-plane surface. The sharp changes at both $T_{CDW}$ and $T^*$ demonstrate the local nature of the "double transition". Reflectivity measurements at various spots are presented in Extended Data Fig.1. Inset: reflectivity near transition temperatures. **(c)** Specific heat capacity $C_p$ exhibits a major peak at $T_{CDW}$ and a minor peak at $T^*$, serving as direct evidence for double phase transitions. Inset: $C_p$ near transition temperatures. **(d)** ScV$_6$Sn$_6$ structure generated using VESTA (left), with the Kagome lattice in the ab-plane (right).

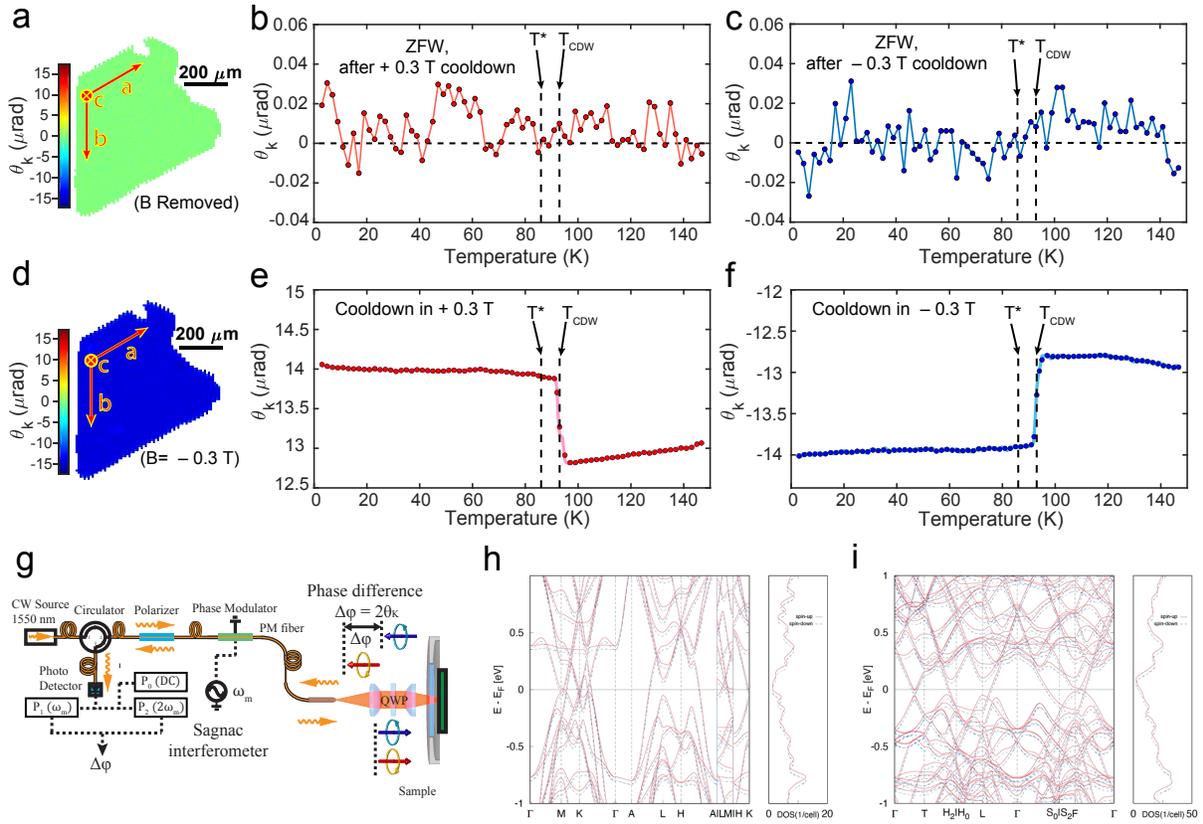

**Figure 2. MOKE $\theta_K$ in ScV$_6$Sn$_6$ and absence of spontaneous time-reversal symmetry breaking (TRSB).** MOKE signal $\theta_K$ is measured using a Sagnac interferometer microscope, first during cooldown recording $\theta_K$ in a magnetic field, then during zero field warmup (ZFW) with the field removed to record spontaneous $\theta_K$. See Methods for details. **(a)** Image of spontaneous $\theta_K$ at $3\,K$ after the removal of a $B = -0.3\,T$ magnetic field, showing zero signal across the sample. **(b), (c)** ZFW after $+0.3\,T$ and $-0.3\,T$ field cooldowns show no discernable onset of spontaneous $\theta_K$ at either $T_{CDW}$ or $T^*$, with $\pm 10\,nrad$ uncertainty. Together these spontaneous $\theta_K$ traces and image show no evidence for TRSB. More measurements of spontaneous $\theta_K$ are presented in Extended Data Fig.3. **(d)** Image of $\theta_K$ at $3\,K$ in $B = -0.3\,T$ showing uniform $\theta_K$ across the sample due to Pauli paramagnetism. **(e), (f)** $\theta_K$ during $+0.3\,T$ and $-0.3\,T$ field cooldowns showing a paramagnetic MOKE response with a sharp increase below $T_{CDW}$, but no change across $T^*$, indicating the $T^*$ transition is not coupled to the magnetic degree of freedom. More measurements of $\theta_K$ in magnetic fields are presented in Extended Data Fig.2. **(g)** Schematics of a zero-area-loop fiber-optic interferometer that is only sensitive to TRSB (MOKE $\theta_K$) effects, which is independent of α. The fiber-optic head can be scanned to simultaneously acquire reflection and MOKE images. **(e), (f)** Calculated band structure and density of states of ScV$_6$Sn$_6$ under an external magnetic field

$B$ for (e) the high-temperature phase and (f) the CDW phase. The spin-up and spin-down channels are marked as the red and blue lines, respectively. The spin polarization at the Fermi level is doubled in the CDW phase, accounting for an increase in $\theta_K$ despite reduced density of states.

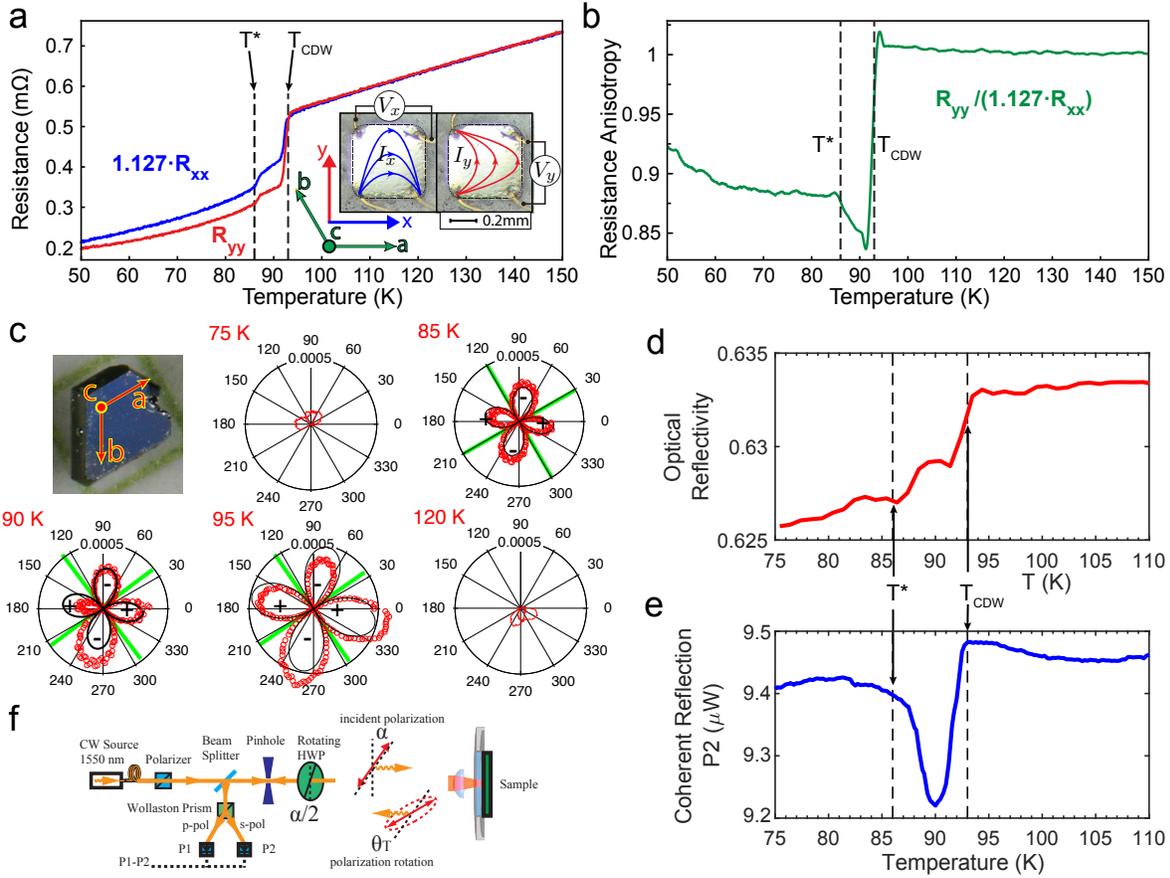

**Figure 3. Broken ab-plane rotational symmetry. (a)** Resistances $R_{xx} \equiv \frac{V_x}{I_x}$ and $R_{yy} \equiv \frac{V_y}{I_y}$ of a ScV$_6$Sn$_6$ sample polished into a square plate. $1.127R_{xx}$ is used in the plot to account for a slight deviation from a perfect square-shape and misalignment of electrical contacts. $1.127R_{xx}$ and $R_{yy}$ trace each other almost perfectly when cooling down from $150\,K$, but develop a pronounced anisotropy below $T_{CDW}$. Such anisotropy decreases but doesn't fully disappear when further cooling down below $T^*$. Inset: pictures of the square sample with illustrations of a-b and x-y axes as well as wire configurations. Curved arrows represent the expected current paths for $I_x$ and $I_y$ in the ab-plane. **(b)** Temperature trace of $\frac{R_{yy}}{1.127R_{xx}}$ as a phenomenological parameter quantifying the ab-plane transport anisotropy. The largest anisotropy occurs between $T_{CDW}$ or $T^*$. **(c)** Polar plots of optical polarization rotation measured at one location on the ab-plane and at various temperatures. The development of a four-leaf clover pattern with alternating "+" and "-" signs between $95\,K$ and $85\,K$ indicates *local* ab-plane anisotropy. The full scale of the polar plots is $0.0005\,rad$. The crossed green lines represent fitted principal birefringent axes, which are close to the a-axis and its normal direction. Inset: image of the measured ScV$_6$Sn$_6$ single crystal with crystal axes labeled. Polarization rotation measurements at various locations are presented in Extended Data Fig.4. **(d), (e)** Depolarization effects revealed by comparing optical reflectivity (red) and the coherent reflection $P_2$

measured in the Sagnac interferometer (blue). $P_2$ measures the part of reflected optical power that remains coherent after reflection in a Sagnac interferometer, where depolarization effects in the sample such as anisotropy and chirality reduce such coherence. See Methods for details. Profoundly, cooling from $T_{CDW}$ to $T^*$, the reflectivity (red) shows a drop-plateau-drop while $P_2$ exabits a single pronounced dip, revealing a large depolarization effect between $T_{CDW}$ and $T^*$, which is consistent with the observed transport and optical anisotropy. Cooling further below $T^*$, $P_2$ recovers from the dip and starts gradually to follow the trend of reflectivity, signaling the weakening and possible disappearance of the ab-plane anisotropy. $P_2$ measurements at various locations are presented in Extended Data Fig.5. **(f)** Schematics of polarization rotation setup based on a Wollaston prism that measures polarization rotation $\theta_T$ as a function of incident polarization $\alpha$, which is achieved by rotating a half-wave plate (HWP) by $\frac{\alpha}{2}$.

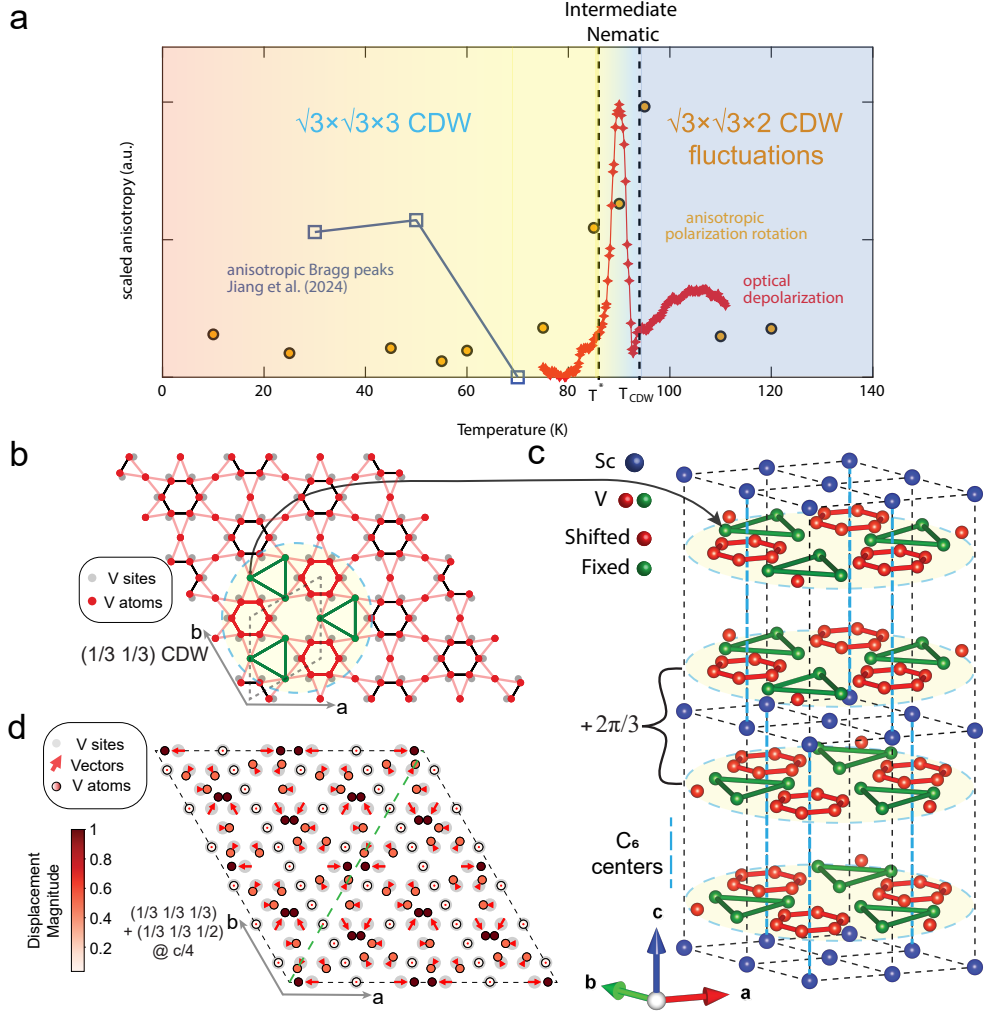

**Figure 4. Summary of nematic phases in ScV$_6$Sn$_6$ and potential sources of the intermediate nematic phase. (a)** Diagram of previously reported intra-unit cell nematic order below 70 $K$ [9] and the intermediate nematic state reported here. Anisotropic polarization rotation (yellow) represents fitted $\theta_P$ values from Fig.3c using $\theta_T = \theta_P \sin(2\alpha - A)$, and optical depolarization (red) is calculated as the coherent reflection $P_2$ (Fig.3d) divided by reflectivity (Fig.3e), both indicating an intermediate anisotropic phase between $T_{CDW}$ and $T^*$ where $\sqrt{3} \times \sqrt{3} \times 3$ and $\sqrt{3} \times \sqrt{3} \times 2$ CDWs compete. Anisotropic Bragg peaks (blue) is calculated as the ratio between Bragg peak intensities at $q_3$ and $q_2$ adopted from a recent STM study [9] suggesting an intra-unit cell nematic order below 70 $K$. **(b)** Illustration of the kagome lattice in the CDW phase with an exaggerated in-plane $\sqrt{3} \times \sqrt{3}$ CDW modulation for Vanadium atoms with wave vector $\left(\frac{1}{3}, \frac{1}{3}\right)$, and 3 times larger in-plane unit cell (dashed black lines). Within the dashed blue circle, displaced and fixed atoms are colored red and green. **(c)** Stacking of CDW states depicted in (b) with $\frac{2\pi}{3}$ interlayer phase shift breaks rotational symmetry. The accumulated $\frac{2\pi}{3}$ phase difference between the top and bottom

layer results in breaking of the $C_6$ symmetry due to a misalignment of the $C_6$ centers. **(d)** Mixing of competing $\sqrt{3} \times \sqrt{3} \times 3$ CDW with $q_s = (\frac{1}{3}, \frac{1}{3}, \frac{1}{3})$ and $\sqrt{3} \times \sqrt{3} \times 2$ CDW with $q^* = (\frac{1}{3}, \frac{1}{3}, \frac{1}{2})$, plotted at height $z = c/4$. The result is an anisotropic modulated lattice marked by the dashed green line. Original vanadium lattice sites of the kagome lattice are marked by gray circles, and displaced vanadium atoms are marked by circles with black edges with inside colors ranging from white (zero displacement) to red (maximum displacement). The darkest red on the color bar is normalized to the largest displacement magnitude. Red arrows at each gray circle represent the displacement vectors from the original lattice site to the displaced atom.

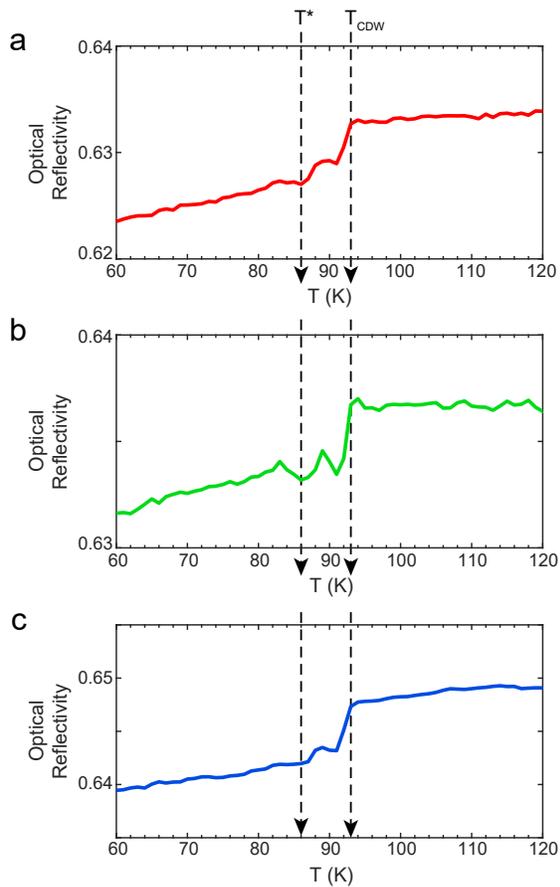

**Extended Data Figure 1. Temperature-dependent optical reflectivity (1.55 $\mu m$) measured at different locations.** Consistently at all measured locations, we observe a ~0.8% drop in optical reflectivity when the sample is cooled through $T_{CDW}$, and another ~0.3% drop when cooling through $T^*$. These findings demonstrate that the observed double phase transitions at $T_{CDW}$ and $T^*$ are not attributable to macroscopically separated volumes in the sample with two distinct CDW transition temperatures, but rather are localized phenomena. The absolute values of optical reflectivity exhibit a spatial variation of 0.015, indicating a small 2% inhomogeneity across the entire sample.

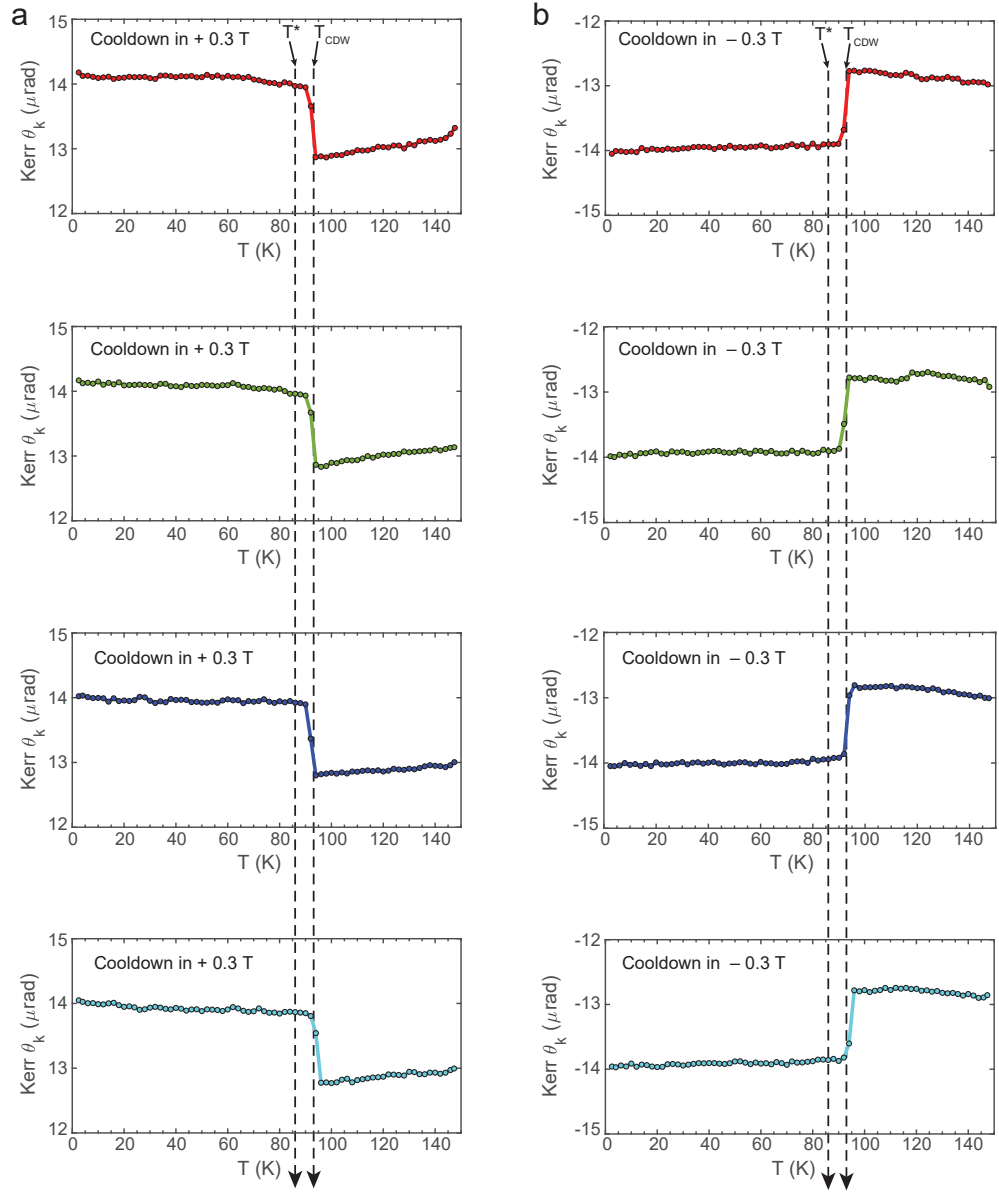

**Extended Data Figure 2. MOKE $\theta_K$ measured during field cooldowns. (a)** $\theta_K$ during B = + 0.3 T field cooldowns. **(b)** $\theta_K$ during B = − 0.3 T field cooldowns. The results show a paramagnetic MOKE response with a sharp increase below $T_{CDW}$, but no change across $T^*$, indicating the $T^*$ transition is not coupled to the magnetic degree of freedom.

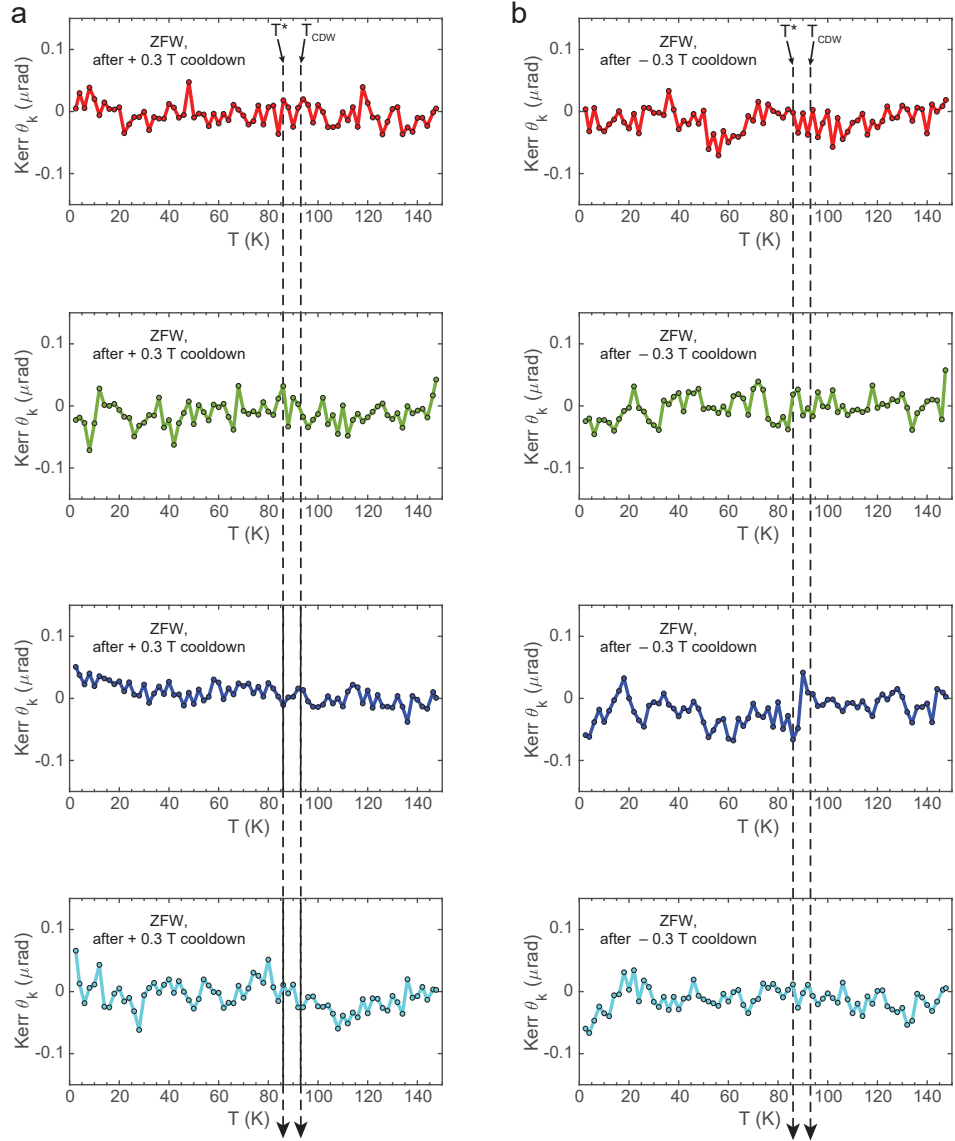

**Extended Data Figure 3. Spontaneous MOKE $\theta_K$ measured during zero-field warmups (ZFW). (a)** *ZFW* after B = + 0.3 *T* field cooldowns. **(b)** *ZFW* after B = − 0.3 *T* field cooldowns. No discernable onset of spontaneous $\theta_K$ was found at either $T_{CDW}$ or $T^*$ with ±30 $nrad$ uncertainty, showing no evidence for TRSB.

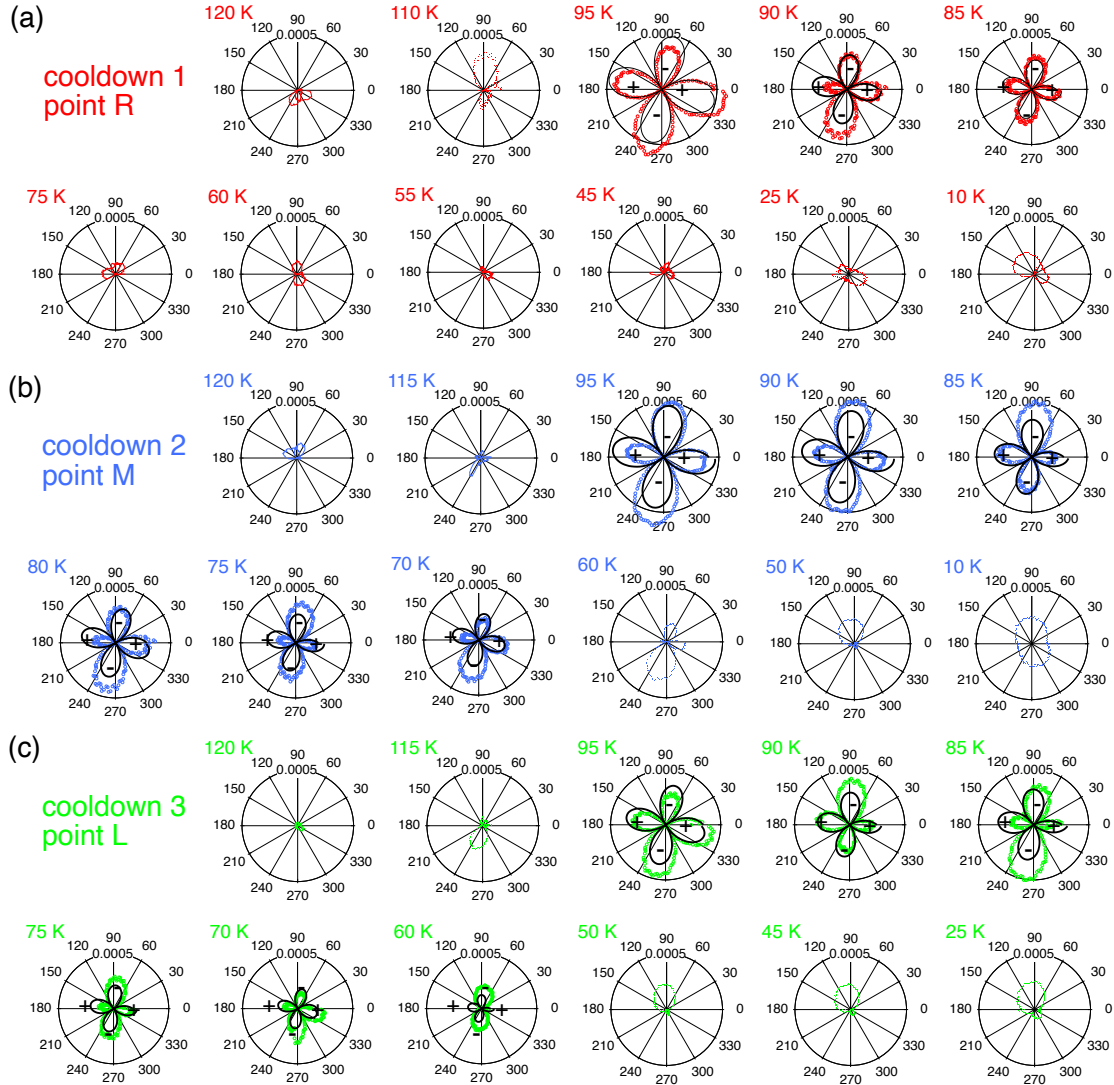

**Extended Data Figure 4. Polar plots of optical polarization rotation measured at various locations on the ab-plane and at various temperatures.** The development of four-leaf clover pattern with alternating "+" and "-" signs indicates *local* ab-plane anisotropy. The full scale of the polar plots is $0.0005\ rad$. While most of the anisotropy is observed between $T_{CDW}$ and $T^*$, at points M and L the anisotropy persists to $70\ K$ and $60\ K$ respectively.

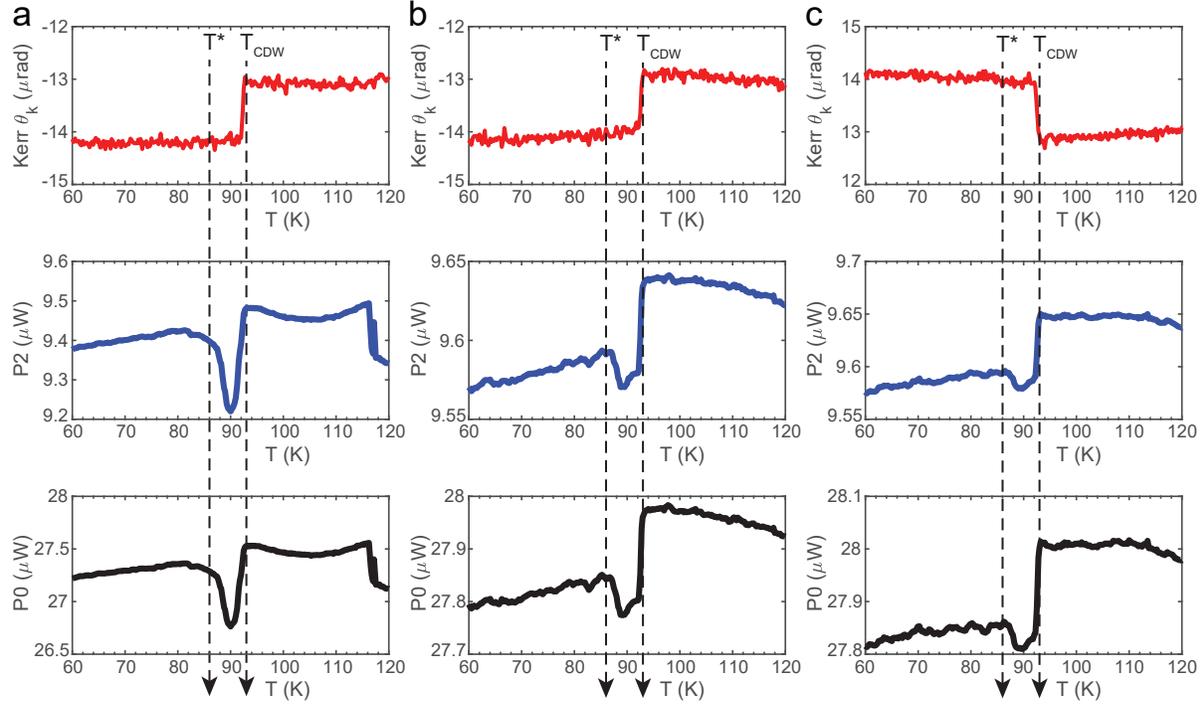

**Extended Data Figure 5. $\theta_K$, $P_2$, and $P_0$ simultaneously recorded in Sagnac interferometer measurements at various locations. (a), (b)** $B = -0.3T$ cooldown. **(c)** $B = +0.3T$ cooldown. As explained in Methods, the measured $\theta_K$ is the MOKE signal, which is independent of sample reflectivity or depolarization effects. $P_2$ and $P_0$ on the other hand record the part of reflected optical power that is not depolarized by the sample surface. The depolarization effect here arises from the ab-plane anisotropy in ScV$_6$Sn$_6$. Profoundly, cooling from $T_{CDW}$ to $T^*$, the reflectivity (Extended Data Fig.1) shows a drop-plateau-drop while $P_2$ and $P_0$ exabit a single pronounced dip, revealing a large depolarization effect between $T_{CDW}$ and $T^*$, which is consistent with the observed transport and optical anisotropy. Further, between different locations, we observe a difference in the size and the lower temperature bound of the dip in $P_2$ and $P_0$. This inhomogeneity echoes that observed in the polarization rotation measurements presented in Extended Data Fig.4.